\documentclass{Interspeech2024}

\interspeechcameraready

\title{Asynchronous Voice Anonymization Using Adversarial Perturbation On Speaker Embedding\thanks{*: corresponding author. This work was supported in part by the National Natural Science Foundation of China under Grant U23B2053.}}
\name[affiliation={1}]{Rui}{Wang}
\name[affiliation={1*}]{Liping}{Chen}
\name[affiliation={2}]{Kong Aik}{Lee}
\name[affiliation={1}]{Zhen-Hua}{Ling}

\address{
  $^1$NERC-SLIP, University of Science and Technology of China (USTC), China\\
  $^2$Department of Electrical and Electronic Engineering, The Hong Kong Polytechnic University, Hong Kong
 }
\email{wangrui256@mail.ustc.edu.cn, \{lipchen, zhling\}@ustc.edu.cn, kong-aik.lee@polyu.edu.hk}

\usepackage{multirow}
\usepackage{cite}
\usepackage{xurl}

\begin{document}

    \maketitle

    \begin{abstract}

        Voice anonymization has been developed as a technique for preserving privacy by replacing the speaker's voice in a speech signal with that of a pseudo-speaker, thereby obscuring the original voice attributes from machine recognition and human perception. In this paper, we focus on altering the voice attributes against machine recognition while retaining human perception. We referred to this as the \emph{asynchronous voice anonymization}. To this end, a speech generation framework incorporating a speaker disentanglement mechanism is employed to generate the anonymized speech. The speaker attributes are altered through adversarial perturbation applied on the speaker embedding, while human perception is preserved by controlling the intensity of perturbation. Experiments conducted on the LibriSpeech dataset showed that the speaker attributes were obscured with their human perception preserved for 60.71\% of the processed utterances. Audio samples can be found in \footnote{https://voiceprivacy.github.io/asynchronous-voice-anonymization}.


    \end{abstract}

    

    \section{Introduction}
        With the rapid development of speech technology in recent years, the security of speech data is confronted with increasing risks. For example, the advances in speaker recognition techniques facilitate recognizing the speaker's identity in a speech utterance with high accuracy. 
        When combined with speech recognition, the personal information conveyed in the speech utterances might be revealed and leaked. Also, the advancements in personalized speech generation techniques, including \emph{voice conversion} (VC)\cite{sisman2020overview,hsu2016voice,qian2019autovc} and \emph{text-to-speech} (TTS)\cite{ning2019review,li2019neural,kim2020glow}, enable the generation of artificial speech utterances in both high quality and speaker similarity. Such techniques can be maliciously used to generate fake speech imitating specific individuals, thereby impersonating their identities. 

        The aforementioned risks of speaker information raise the call for voice privacy protection techniques. In this regard, the anti-spoofing\cite{wu2015spoofing} methods were developed to detect artificially generated speech utterances. They operate after the speaker information is extracted and applied to generate the fabricated speech. By preventing the fabricated speech from being used for malicious purposes, the anti-spoofing methods protect the speaker information in a passive manner. On the other hand, the voice anonymization technology \cite{2020Introducing,mcadams1984spectral,fang2019speaker,srivastava2022privacy} protects the speaker information within a speech utterance by replacing the original speaker with a pseudo-speaker by means of VC. It provides proactive protection to the information of the original speaker once it's generated. Such a speaker replacement approach alters not only the speaker information perceived by machine learning algorithms but also that perceived by human listeners.
        

        However, application scenarios also call for the voice privacy protection under the condition that the human perception of the original speaker is preserved, e.g., the famous people who address the public, the content creators of YouTube short videos, etc. In this paper, we focus on anonymizing the voice when perceived by machine algorithms while preserving human perception. In contrast to the existing voice anonymization where both machine and human perceptions of speaker attributes are altered synchronously, the proposed task is asynchronous. To this end, a speech generation framework is adopted to incorporate an explicit mechanism for speaker information modeling. Within this framework, the speaker representation extracted from the original speech is modified and used to generate the protected speech. In our work, the VC function of the YourTTS\cite{casanova2022yourtts} model is utilized given its capability in information disentanglement and speech generation. As a preliminary work, this paper focuses on protecting closed-set speakers. The \emph{fast gradient step method} (FGSM) algorithm \cite{2014Explaining} is adopted to modify the speaker embedding. Experiments were conducted on the LibriSpeech dataset\cite{librispeech}. Listening test reveals that 60.71\% of the processed utterances sounded alike to those utilizing original speaker attributes. In automatic speaker verification evaluations, higher equal error rates (EERs) of these utterances suggest their capability to obscure speaker attributes for machines while maintaining them to humans.

    The rest of the paper is organized as follows.
    Section \ref{sec: background} describes the background of our work.
    The proposed protected speech generation method is illustrated in Section \ref{sec: protected speech generation}.
    Experimental results and analysis are presented in Section \ref{sec: experiments}.
    Finally, Section \ref{sec: conclusions} concludes the paper.

    \section{Background}
    \label{sec: background}
    In this section, we will first briefly go through the speaker embedding modeling. Then, the inference flow of the VC function within YourTTS model is described.
    Finally, the FGSM algorithm for adversarial attack is presented.
    

   \begin{figure}[t]
        \centering
        \includegraphics[scale=0.4]{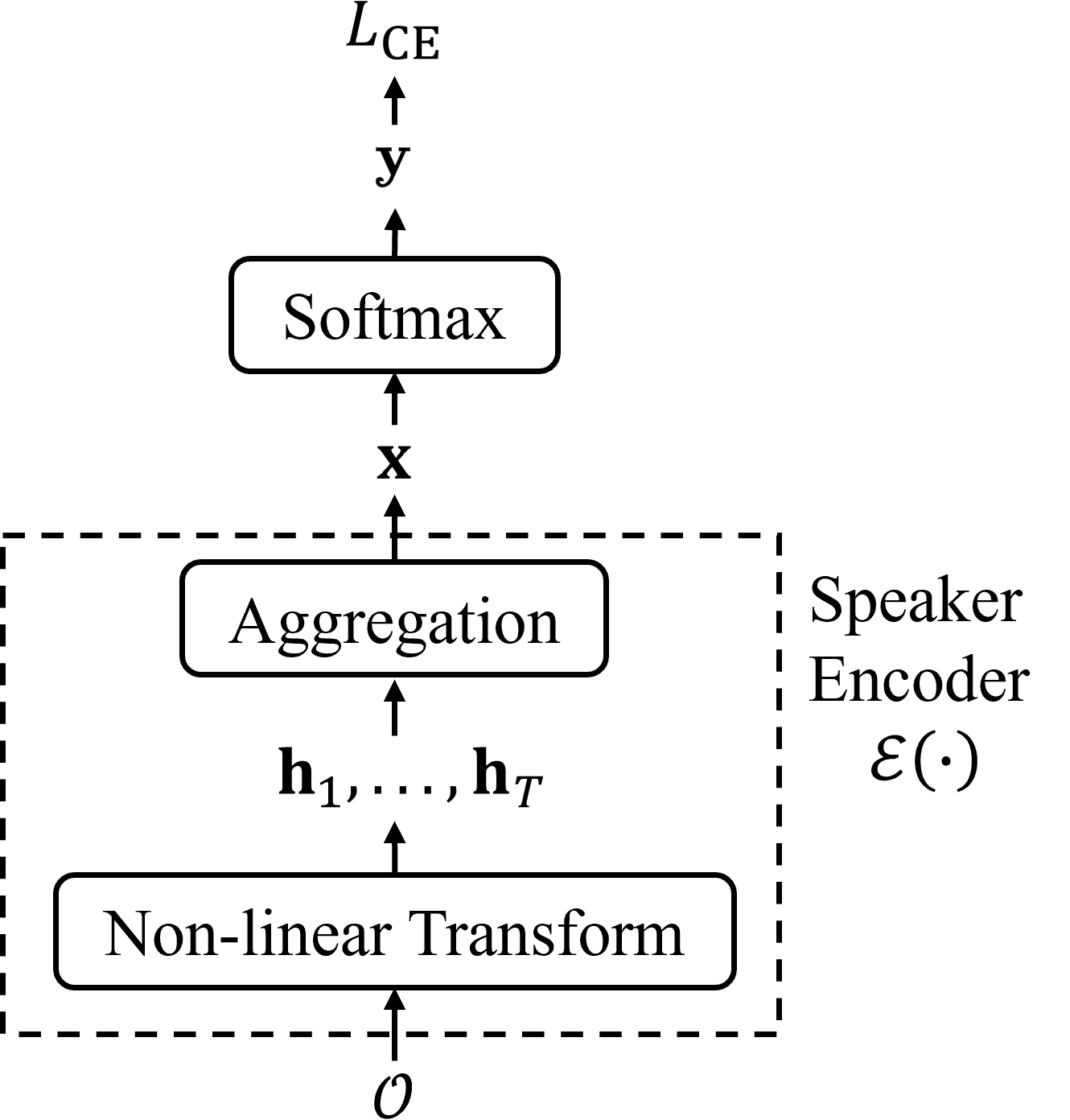}
        \caption{The architecture of the discriminative speaker
attributes modeling. The input
speech utterance is represented as $\mathcal{O}$. Its speaker embedding vector $\mathbf{x}$ is extracted with the speaker encoder $\mathcal{E}\left(\bullet\right)$.}
        \label{fig:Speaker Embedding}
    \end{figure}

    \subsection{Speaker embedding}
    Speaker embedding refers to the representation of a variable-length utterances as a fixed-length vector\cite{lee2021xi}.
    Mathematically, denoting an utterance as $\mathcal{O}=\left\{{\bf o}_{1},...,{\bf o}_T\right\}$ with $T$ denote the number of frames, a speaker embedding is estimated as:
    \begin{equation}
    \label{eq. speaker encoder}
        {\bf x}=\mathcal{E}\left({\bf o}_{1},...,{\bf o}_T\right),
    \end{equation}
    where $\mathcal{E}\left(\bullet\right)$ represents the speaker embedding extractor.
    
    In the neural network framework, $\mathcal{E}\left(\bullet\right)$ is always specified to be the speaker encoder, trained in a discriminative manner. Fig. \ref{fig:Speaker Embedding} describes the neural network architecture for speaker embedding modeling.
    As illustrated in Fig. \ref{fig:Speaker Embedding}, $\mathcal{E}\left(\bullet\right)$ firstly performs non-linear transforms on $\mathcal O$ and outputs $\left\{{\bf h}_{1},...,{\bf h}_T\right\}$ as the representations of the speaker attributes within each frame. Then it aggregates the frame-level embedding vectors into an utterance-level embedding vector ${\bf x}$. Following that, a softmax layer is applied on ${\bf x}$ to predict the speaker class of $\mathcal O$ as follows:
    \begin{equation}
        \label{eq. NN prediction}
        {\bf y}=softmax\left({\bf x}\right),
    \end{equation}
    with $\bf y$ denoting the output of the softmax layer.
    
    The cross-entropy (CE) between $\bf y$ and the label ${\bf t}$ is computed as the loss function as follows:
    \begin{equation}
        \label{eq. CE loss}
        L_{\rm CE}=-\sum_{c=1}^C{t}_c{\rm log}{y}_c,
    \end{equation}
    with $C$ denoting the number of speaker classes. As such, the model is trained by minimizing $L_{\rm CE}$. The speaker encoder $\mathcal{E}\left(\bullet\right)$ can be obtained as a result. In inference, given a speech utterance, the speaker encoder is applied to estimate the vector $\bf x$, referred to as the speaker embedding.


    \begin{figure}[!h]
        \centering
        \includegraphics[scale=0.4]{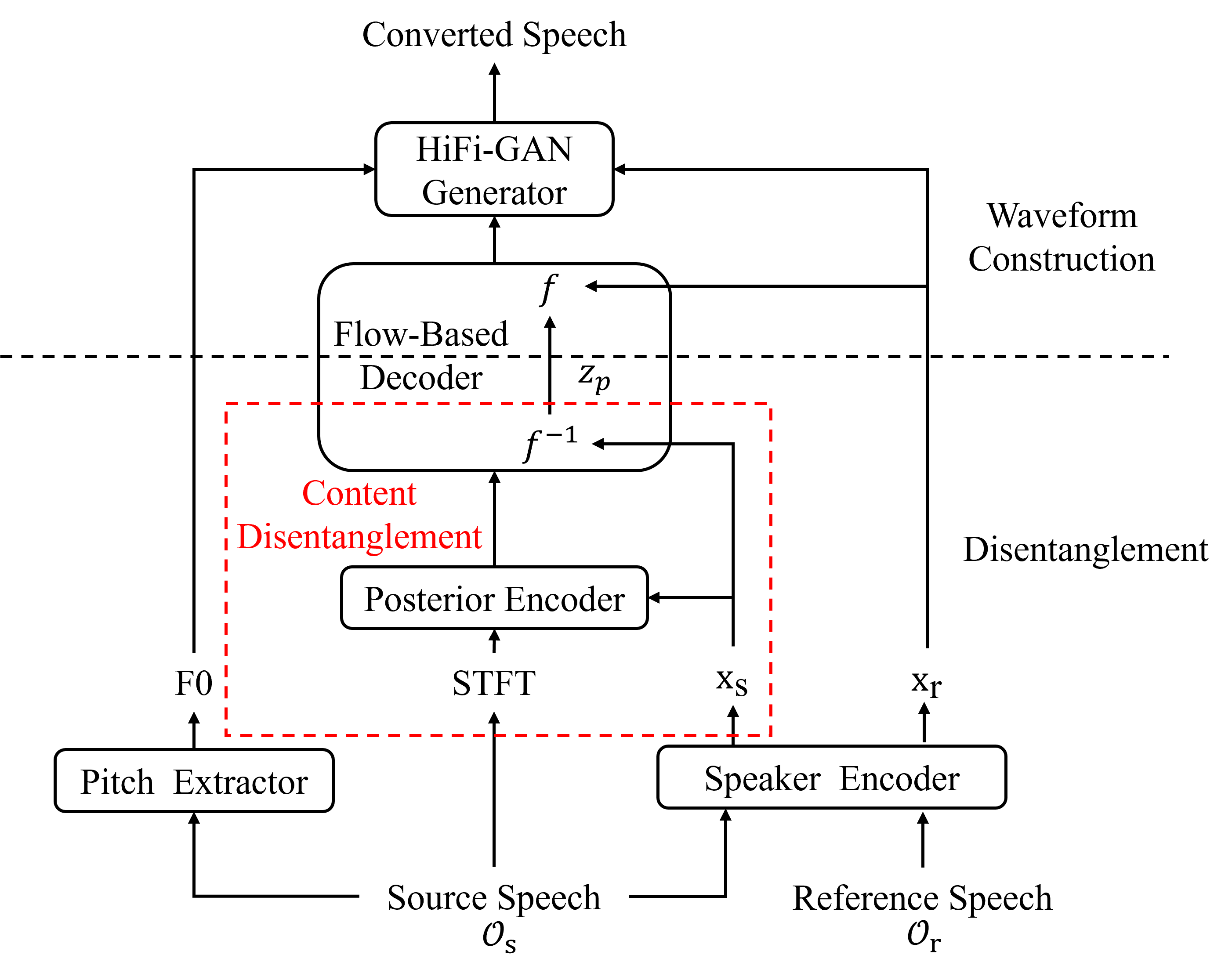}
        \caption{Inference flow of the voice conversion function of YourTTS. In comparison with the official version as described in \cite{casanova2022yourtts}, the F0 extracted from the source speech is used. The black dotted line separates the modules into information disentanglement and waveform construction. The modules in the rectangular box of red dotted line are used for disentangling the content information from the source speech.}
        \label{fig:YourTTS}
    \end{figure}

    \subsection{YourTTS}
    YourTTS is a model used for personalized speech generation in both TTS and VC functions\cite{casanova2022yourtts}.
    The VC function is adopted in our work, which converts the speaker in the source speech to that of the reference speech. The inference flow of the VC function can be decomposed into information disentanglement and waveform construction as illustrated in Fig. \ref{fig:YourTTS}.
    
    In detail, the model takes a pair of source and reference speech utterances as input, denoted as $\mathcal{O}_{\rm s}$ and $\mathcal{O}_{\rm r}$, respectively. In the disentanglement stage, the content information in $\mathcal{O}_{\rm s}$ is disentangled via the content disentanglement module. Specifically, this is fulfilled by the posterior encoder\cite{kim2021conditional} and the reverse pass of the flow-based decoder $f^{-1}$. The content information is represented as ${\bf z}_p$ in Fig. \ref{fig:YourTTS}. In this process, the speaker embedding vector ${\bf x}_{\rm s}$ is extracted from ${\mathcal O}_{\rm s}$ first and then used to remove the speaker information from the content representation. Meanwhile, the speaker information within ${\mathcal O}_{\rm r}$ is extracted via the speaker encoder, represented with the speaker embedding vector ${\bf x}_{\rm r}$. In our work, the F0 values of the source speech ${\mathcal O}_{\rm s}$ are extracted with a pitch extractor, representing its prosody information.    
    Finally, the prosody and content representation of ${\mathcal O}_{\rm s}$ combined with the speaker representation of ${\mathcal O}_{\rm r}$ to construct the waveform. This is achieved via the forward pass of the flow-based decoder followed by the HiFi-GAN\cite{kong2020hifi} generator. Finally, the speech is generated by converting the speaker attributes within the source speech $\mathcal{O}_{\rm s}$ into that of the reference speech ${\mathcal{O}}_{\rm r}$. For details, readers are suggested to read \cite{casanova2022yourtts}.


    \subsection{FGSM}
    The \emph{fast gradient step method} (FGSM) was proposed in \cite{2014Explaining} to generate adversarial samples to attack neural network models.
    Given an input sample {\bf s} and the network model ${\bf y}=f_{\theta}({\bf s})$ where $f_{\theta}(\bullet)$ is the network function with the subscript $_{\theta}$ denoting the model parameter set.
    Denote the loss function as $L({\bf y}, {\bf y}^{{\rm true}})$, where ${\bf y}^{\rm{true}}$ is the true label associated with {\bf s}.
    The adversarial sample is obtained by taking a small step in the direction that maximizes the loss function with respect to $\bf s$. Mathematically, $\bar{{\bf s}}$ is obtained as:
    \begin{equation}
    \label{eq. FGSM}
        \bar{\bf s}={\bf s}+\epsilon \cdot sign\left(\nabla_{\bf s} L\left(f_{\theta}\left(\mathbf{s}\right),\mathbf{y}^\mathrm{true};\theta\right)\right).
    \end{equation}
    \begin{figure}[!t]
        \centering
        \includegraphics[scale=0.38]{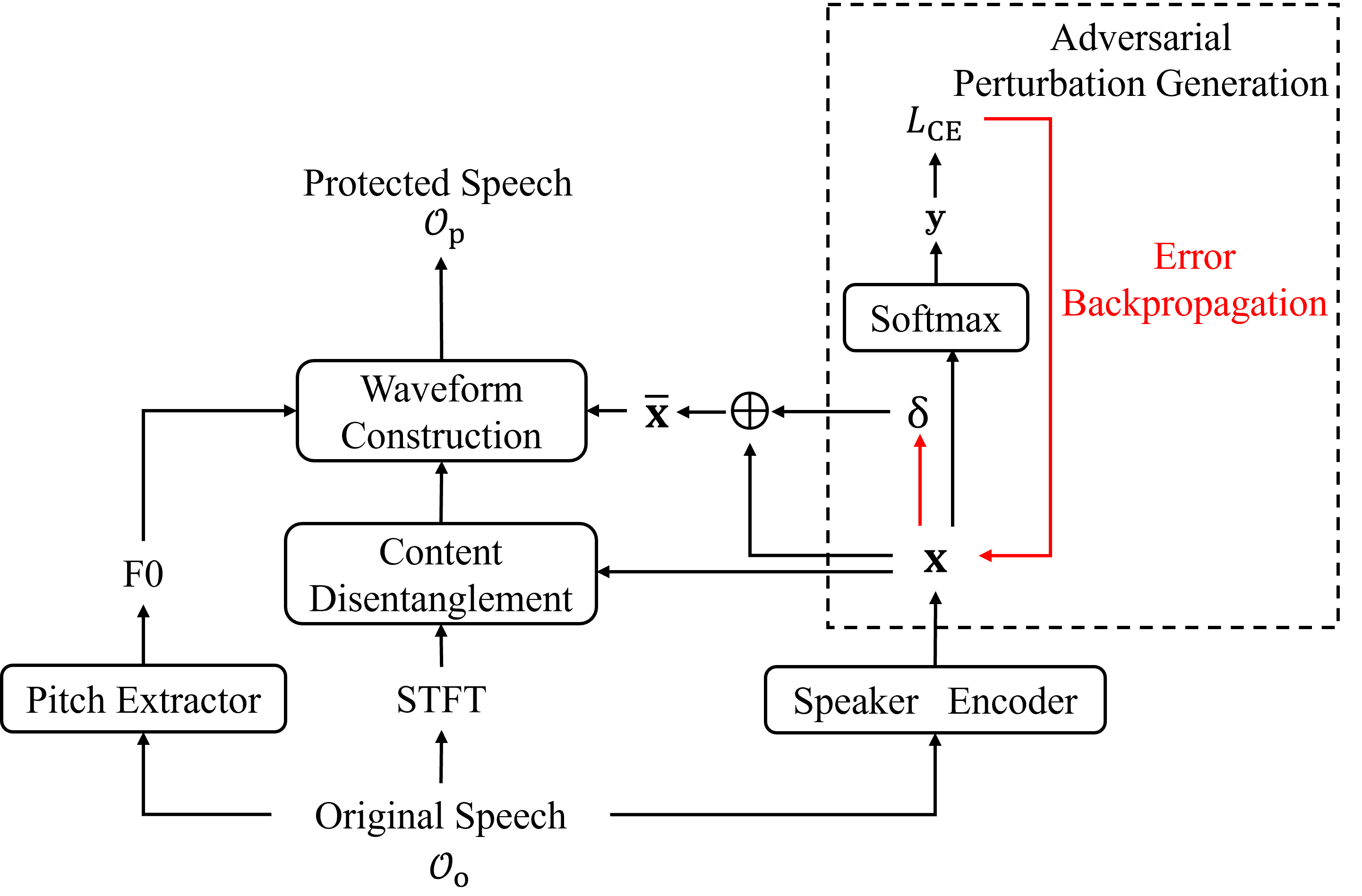}
        \caption{Protected speech generation based on the VC function of YourTTS. The content disentanglement module inherits that from Fig. \ref{fig:YourTTS}. The rectangular box of the dotted line is the adversarial attack to generate the perturbed speaker embedding ${\bar {\bf x}}$. The error backpropagation process to obtain the perturbation $\delta$ is denoted by the red arrow line.}
        \label{fig:perturbed_embedding}
    \end{figure}
    In (\ref{eq. FGSM}), $\epsilon$ is a variable controlling the intensity of the attack, $sign\left(\bullet\right)$ is the operation that takes the sign of the value, $\nabla_{\bf s}$ is the derivative with respect to the original sample $\mathbf{s}$.

    \section{Protected speech generation} 
    \label{sec: protected speech generation}
    This section describes the proposed method for speaker protected speech generation, including the overall model architecture and the adversarial perturbation on the speaker embedding.

    \subsection{Overall architecture}
    Fig. \ref{fig:perturbed_embedding} depicts the diagram of the generation of the protected speech. In detail, given the VC flow of a YourTTS model, the original speech utterance ${\mathcal O}_{\rm o}$ is used as both the source and reference utterances. The original speaker embedding vector ${\bf x}$ is firstly extracted from ${\mathcal O}_{\rm o}$. The content information is disentangled from ${\mathcal O}_{\rm o}$ via the content disentanglement module with the help from ${\bf x}$, and denoted with ${\bf z}_p$. Meanwhile, the F0 values are extracted, representing the prosody information in ${\mathcal O}_{\rm o}$. To generate the protected speech $\mathcal{O}_{\rm p}$, the speaker embedding vector ${\bf x}$ is modified to be ${\bar {\bf x}}$. In our work, the modification is accomplished by the adversarial perturbation generation module as illustrated in the rectangular box of the dotted line. To be specific, the perturbation is generated through the adversarial attack and is described in the following.
    
    \subsection{Adversarial perturbation on speaker embedding}
    Given the speaker model within which the speaker encoder is trained, it is perturbed using the FGSM algorithm. First of all, assume that the speaker of ${\mathcal O}_{\rm o}$ is among the speaker classes that the speaker model can recognize, its speaker can be predicted with the softmax function on ${\bf x}$ as follows:

    \begin{equation}
        \label{eq. softmax layer}
        {\bf y}=softmax({\bf x}),
    \end{equation}
    where $\bf y$ is the prediction output. The cross-entropy between $\bf y$ and the label is computed to be $L_{\rm CE}$ according to (\ref{eq. CE loss}). Following that, as shown in (\ref{eq. FGSM}), the adversarial perturbation $\bm \delta$ is obtained on ${\bf {\rm x}}$ via error backpropagation. The adversarial speaker embedding vector ${\bar {\bf x}}$ is obtained by adding $\bm \delta$ to ${\bf x}$ as follows:
    
    \begin{equation}
    \label{eq. adversarial speaker embedding}
        \bar{\bf x}={\bf x}+\epsilon\cdot{\bm \delta},
    \end{equation}
    with $\epsilon$ denotes the intensity.
    The protected speech waveform ${\mathcal O}_{\rm p}$ is constructed by combining the F0 values, the speech content representation ${\bf z}_p$ and the adversarial speaker embedding ${\bar{\bf x}}$.

     \section{Experiments}
     \label{sec: experiments}

    \subsection{Dataset \& configurations}
   Our experiments were conducted on the LibriSpeech corpus. Within which, the dev-clean and test-clean datasets were used for evaluation, including 5,323 utterances from 40 female and 40 male speakers. 
   In our experiments, YourTTS model was trained on the LibriTTS\cite{zen2019libritts} train-clean-100 dataset. 
   The speaker encoder utilized the pre-trained H/ASP\cite{heo2020clova} model available in the open-source repository\footnote{https://github.com/clovaai/voxceleb\_trainer}.
    The open-source H/ASP speaker encoder was trained on the VoxCeleb2\cite{chung18b_interspeech}. However, the FGSM algorithm in our protected speech generation requires that the original speaker must be one of the training speakers for the speaker encoder. To this end, a softmax layer is retrained on the speaker encoder with utterances from the test speakers. 
    For each test speaker, 70\% utterances were selected to train a softmax layer. 
    In this process, only the parameters of the softmax layer were trained on the speaker encoder, while the remaining parameters of the speaker encoder were kept frozen.
     The rest 1,522 utterances of the speakers were used in our evaluations. In the FGSM algorithm implemented on the speaker embedding during protected speech generation, the intensity $\epsilon$ was set to 0.02.

    
    In the evaluations, three kinds of speech utterances were evaluated, including: 
    1) the original speech;
    2) the regenerated speech that the YourTTS model generated without modifying the speaker embedding vector;
    3) the protected speech generated by adding perturbation to speaker embedding. Initially, a subjective listening test was applied to assess the ability of protected utterances to maintain human perception of speaker attributes. Given the precondition for the protected utterance to preserve the human perception of speaker attributes, utterances maintaining the human perception of the speaker attributes were chosen for subsequent evaluations. Next, a \emph{speaker mean opinion score} (SMOS) test was conducted further to evaluate the preservation of speaker attributes in human perception. Following this, ASV evaluations were applied to measure speaker identity concealment from machine algorithms. Finally, ASR evaluations were conducted to measure the preservation of content information in the protected speech.

    \subsection{Human perception evaluation}
    First of all, a subjective listening test was conducted on the protected utterances to assess their ability to maintain the human perception of the speaker attributes. As the protected speech was generated using the YourTTS model, potentially introducing quality reduction compared to the original ones, regenerated utterances were used as the reference for a fair speaker comparison. In this test, 1,522 protected utterances were generated from the rest 1,522 utterances, three listeners were asked to decide whether the protected utterances sounded alike to the corresponding regenerated ones. The listeners provided binary responses (yes or no) for each pair. The utterance pairs that obtained yes from all three listeners were decided to have the same speaker attributes.
    
    Through listening test, 60.71$\%$ of the protected speech utterances were labeled to match the speakers within the regenerated utterances. This shows that protected speech is capable of preserving the human perception of the speaker attributes. These utterances were selected and used in our remaining evaluations, called the labeled dataset, including 455 utterances from 36 female and 469 utterances from 37 male.

    \subsection{SMOS test}
    The \emph{speaker-similarity mean option score} (SMOS) test was conducted by 15 paid native speakers. They were asked to score the speaker similarity between utterance pairs.
    SMOS scores ranged from 1 to 5 with a 0.5 step.
    The higher scores indicated higher speaker similarity.
    The SMOS tests were carried out in three scenarios, where the utterance pair were: \emph{original-original(ori-ori)}, \emph{original-regenerated(ori-regen)} and \emph{regenerated-protected(regen-prote)}.
    The SMOS scores are presented in Table \ref{tab:1}.
    \begin{table}[!htbp]
        \caption{The SMOS scores with 95$\%$ confidence intervals among the original(ori), regenerated(regen), and protected(prote) utterances generated with out proposed method.}
        \label{tab:1}
        \centering
        \begin{tabular}{ccccc}
            \toprule
            & ori-ori  & ori-regen & regen-prote \\
            \midrule
            SMOS & $4.17\pm0.14$      & $4.08\pm0.10 $      & $4.16\pm0.10$         \\
            \bottomrule
        \end{tabular}
    \end{table}
    
    As shown in Table \ref{tab:1}, by comparing the score of ori-ori and ori-regen, it can be seen that the speech generation of our YourTTS model did not significantly affect the speaker attributes within the original speech utterances in terms of human hearing. Furthermore, the protected utterances were compared to the regenerated utterances, instead of the original ones, to mitigate the speaker similarity degradation caused by the YourTTS model, ensuring a fair evaluation. The SMOS of regen-prote was close to ori-ori, which serves as the benchmark for SMOS. This indicates the protected utterances sounded alike to humans with the regenerated utterances and the protected speech was able to preserve the speaker attributes for humans.

    \subsection{ASV evaluations}
    To evaluate the capability of the protected speech in protecting the speaker privacy, the ASV evaluations were carried out with the performances being measured by the \emph{equal error rate} (EER).
    The ASV tests were conducted on the x-vector\cite{x-vector} model, i-vector\cite{i-vector} models, and the speaker encoder of the YourTTS model. 
    According to whether the evaluation model was seen during speech generation, the first two models were black-box while the third one is white-box.
    The x-vector and i-vector models were trained following the Kaldi recipe\footnote{https://github.com/kaldi-asr/kaldi/egs/voxceleb}\cite{povey2011kaldi} using the original speech and the regenerated speech of train-clean-360 dataset.
    The scores were computed with a \emph{probabilistic linear discriminant analysis }(PLDA)\cite{prince2012computer} backend. The ASV evaluations were conducted in a gender-dependent manner on the 60.71\% test utterances within which the speaker attributes sounded similar between the regenerated and the protected utterances. In the evaluation trial configuration, 25$\%$ of the test utterances of each speaker were used for enrollment with the other 75$\%$ used for test. The target and nontarget trials were composed in a ratio of 1:35 for each speaker. Overall, the evaluation on females involved 3,003 target and 107,158 nontarget trials, while the evaluation on males included 2,559 target and 91,463 nontarget trials. The EERs are given in Table \ref{tab:2}.

    \begin{table}[h]
        \caption{The EERs($\%$) calculated on the original(ori), regenerated(regen), and protected(prote) utterances generated with our proposed method. Evaluation results on the speaker encoder within the YourTTS model, x-vector, and i-vector are included.}
        \label{tab:2}
        \centering
        \setlength{\tabcolsep}{0.8mm}{
            \begin{tabular}{cccccc}
                \toprule
                Model                    & training data              & gender & ori & regen & prote \\
                \midrule
                
                \multirow{2}{*}{speaker encoder}    & \multirow{2}{*}{ori} & male   & 0.81      & 4.30      & 9.20     \\
                &                            & female & 0.96      & 4.20     & 9.53     \\
                \midrule
                \multirow{4}{*}{x-vector} & \multirow{2}{*}{ori} & male   & 1.93      & 3.97     & 8.22     \\
                &                            & female & 1.63      & 4.30     & 8.09     \\
                \cmidrule(r){3-6}
                & \multirow{2}{*}{regen} & male   & 3.12      & 3.34      & 7.15      \\
                &                            & female & 3.19      & 3.04      & 6.90     \\
                \midrule
                \multirow{4}{*}{i-vector} & \multirow{2}{*}{ori} & male   & 1.87      & 4.97     & 8.51     \\
                &                            & female & 1.03      & 4.08      & 7.64     \\
                \cmidrule(r){3-6}
                & \multirow{2}{*}{regen} & male   & 3.40      & 3.29      & 6.92     \\
                &                            & female & 1.63     & 2.22      & 5.27     \\
                \bottomrule
            \end{tabular}
        }

    \end{table}

     
Based on the findings in Table \ref{tab:2}, it can be observed that higher EERs were obtained when there was a mismatch in the utterance types between the model training data and the evaluation data, i.e., the original and the regenerated utterances.
This was due to the channel mismatch issue between the two types of utterances. As observed in Table \ref{tab:2}, the protected utterances achieved higher EERs on all evaluation conditions compared to the regenerated utterances. This suggests that within our test dataset, the machine's perception of the speaker attributes in 60.71\% of utterances was obscured, while human perception remained unchanged.

    \subsection{ASR evaluations}
    Finally, in our experiments, the ASR evaluations were conducted on the protected utterances. In this evaluation, the ASR service whisper\footnote{https://platform.openai.com/docs/api-reference/introduction}\cite{radford2023robust} provided by OpenAI was called for speech recognition. The performances were measured with \emph{word error rate} (WER).
    Our evaluations were conducted on the libri-dev and libri-test datasets as officially published within the LibriSpeech corpus. The original, regenerated, and protected utterances were evaluated. The WERs are presented in Table \ref{tab:4}.

    \begin{table}[!htbp]
        \caption{Automatic speech recognition results WERs($\%$) on original(ori), regenerated(regen) and protected(prote) utterances generated with out proposed method.}
        \label{tab:4}
        \centering
        \begin{tabular}{ccccc}
            \toprule
            & ori & regen & prote \\
            \midrule
            WER & 4.26      & 7.47      & 7.66      \\
            \bottomrule
        \end{tabular}
    \end{table}
    
    As shown in Table \ref{tab:4}, Compared with the original utterances, a higher WER was obtained on the regenerated speech.
    This might be caused by that our regenerated utterances were unseen by the ASR model of whisper. Meanwhile, the protected utterances obtained a close WER with the regenerated ones, indicating that the protection on the speaker attributes didn't hurt the content information within the speech utterances.
   
    \section{Conclusions}
    \label{sec: conclusions}
    This paper focuses on the \emph{asynchronous voice anonymization} technique. In our work, a speech generation framework that incorporates the speaker disentanglement mechanism is adopted. The proposed method involves extracting the speaker embedding from the original speech and modifying it with adversarial perturbation to obscure speaker attributes from machine recognition, while human perception is preserved by controlling the magnitude of the perturbation. Experiments conducted on the LibriSpeech dataset showed that the machine recognition of speaker attributes was obscured with a 60.71\% success rate while human perception was preserved. Our future research will focus on improving the modification of speaker embedding to conduct asynchronous voice anonymization on all utterances.

\bibliographystyle{IEEEtran}
\bibliography{mybib}

\end{document}